\documentclass[aps,prb,twocolumn,showpacs,floatfix]{revtex4}

\usepackage{float,graphicx}

\usepackage{psfig,verbatim}

\begin{document}

\title{Hysteresis and noise in ferromagnetic materials with parallel domain walls}

\author{Benedetta Cerruti$^{(1)}$, Gianfranco Durin$^{(2,3)}$ and Stefano Zapperi$^{(4,3)}$}

\affiliation{$^{(1)}$ Departament d'Estructura i Constituents de la Mat\`eria,  Universitat de Barcelona, Mart\' i Franqu\`es 1, Facultat de F\'isica, 08028
  Barcelona, Spain}
\affiliation{$^{(2)}$ Istituto Nazionale di Ricerca Metrologica, strada delle
  Cacce 91, 10135 Torino, Italy}
\affiliation{$^{(3)}$ ISI foundation, Viale S. Severo 65 - 101
33 Torino - Italy}

\affiliation{$^{(4)}$ CNR-INFM National Center on nanoStructures and bioSystems at Surfaces (S3), Dep.of Physics, University of Modena and Reggio Emilia, 
Via Campi 213/A, 41100 Modena - Italy}

\date{\today}

\begin{abstract}

We investigate dynamic hysteresis and Barkhausen noise 
in ferromagnetic materials with a huge number of parallel and rigid Bloch 
domain walls. Considering a disordered ferromagnetic 
system with strong in-plane uniaxial anisotropy and in-plane magnetization driven by an external magnetic field, 
we calculate the equations of motion for a set of coupled domain walls, 
considering the effects of the long-range dipolar interactions and disorder.
We derive analytically an expression for the magnetic susceptivity, related
to the effective demagnetizing factor, and show that it has a logarithmic
dependence on the number of domains.
Next, we simulate the equations of motion and study the effect of the external 
field frequency and the disorder on the hysteresis and
noise properties. The dynamic hysteresis is very well explained by means of
the loss separation theory.

\end{abstract}


\maketitle

\section{introduction}

The study of ferromagnetic hysteresis represents an open field 
of current interest, due to the applications in magnetic recording 
technology and spintronic devices \cite{bertotti}. 
From a purely theoretical point of view, the 
dynamics of disordered magnetic systems represents a central problem in non-equilibrium statistical mechanics. One of the central questions arising in 
the analysis of ferromagnetic systems, is the link between the domain structure
and the hysteretic properties, such as the coercive field, the power losses
and the noise. From the experimental point of view, it is possible to
observe the domain structure on the surface of the sample by magneto-optical
techniques, while the bulk behavior is accessible by inductive techniques.
Several models have been developed to understand the experimental
results, ranging from Ising type models considering the reversal of a set
of interacting spins with \cite{RFIM,sethna,vives,vives2,vives3} or without
disorder \cite{chakrabarti,acharyya,sides,rikvold,rikvold1}, to models focusing
on the dynamics of a single domain wall \cite{lyuksyutov,ABBM,zapperi,zapperi2,urbach,narayan,bahiana,queiroz}.
These two classes of models represent two extreme situations: spin models are
appropriate when dipolar interactions are negligible. On the other hand, in soft magnetic materials
magnetostatic effects induce wide parallel domain walls that determine the magnetization
properties.

The dependence of the hysteresis loop area on the 
frequency and the amplitude of the applied magnetic field is
often referred to as dynamic hysteresis \cite{chakrabarti,moore}
in the context of thin films and to power losses in bulk
materials. In bulk materials, energy dissipation is dominated
by eddy currents propagation \cite{bertotti}, while in thin films this effect
is negligible. This led to the general belief that dynamic
hysteresis in thin films is ruled by completely different laws
than in bulk samples. The results of the models used to 
analyze dynamic hysteresis  are often interpreted by assuming a universal scaling law
for the dependence of hysteresis loop area $A$ on the temperature of the system $T$,
the applied field frequency $\omega$ 
and amplitude $H_0$. The experimental estimates of the scaling 
exponents display a huge variability
\cite{he,luse,jiang,raquet,suen,bland1,choi,lee2,lee3,suen2}, and the
validity of a universal scaling law is still under debate 
\cite{santi,nistor}. 
In this respect, it was recently shown that 
the theory of loss separation developed for bulk materials
could be equally well applied to thin films, since the precise 
nature of the damping (i.e. eddy currents or spin relaxation)  
does not change the basic equations~\cite{colaiori}.

The Barkhausen effect \cite{barkhausen} consists in the irregularity of the magnetization variation while  magnetizing a sample with a slowly varying 
external magnetic field. It is due to the 
jerky motion of the domain walls in a system with structural disorder 
and impurities \cite{BK}.  Once the origin of Barkhausen noise (BN) 
was understood \cite{wisho}, it was soon 
realized that it could be used as a non-destructive and non-invasive
probe to investigate the magnetization dynamics in magnetic materials.
From a theoretical point of view, it is a 
good example of dynamical critical behavior, as evidenced by experimental 
observations of power law distributions for the statistics 
of the avalanche size and duration \cite{sethna}. 
There is a growing evidence that soft magnetic bulk materials can
be grouped into different classes according to the scaling 
exponents values \cite{dz}.
The Barkhausen noise is also an example of dynamics of a system 
presenting collective pinning when a quenched disorder is present,
and it belongs to the family of the so-called crackling noises \cite{sethna}.
This kind of noise is exhibited by a wide variety of physical systems,
from earthquakes on faults to paper crumpling. So the relative ease to
study crackling noise in magnets make the BN 
useful to get a deeper insight on different complex systems.

The most successful models used to describe the BN 
have considered the dynamics of a single domain wall
\cite{lyuksyutov,ABBM,zapperi,zapperi2,urbach,narayan,bahiana,queiroz}, where 
the effect of the other domain walls is accounted for  by an average
demagnetizing field \cite{urbach}.
This simplification is justified because
the Barkhausen noise is usually measured around the coercive 
field where the signal is stationary \cite{durin06}. In this regime, 
the domain walls are reasonably distant from each other
and one can assume that their mutual interaction is negligible.
The validity of this assumption, however, has never been 
established firmly: multi-wall effects may in principle affect
the domain walls dynamics, and thus dynamic hysteresis and BN. 
Even if we consider a single domain wall, its magnetostatic energy 
should depend on the number of the domain walls at least in an effective
medium sense. In particular, the demagnetizing factor, that plays a crucial
role in the BN statistics \cite{zapperi}, should be correctly evaluated
only considering the entire domain structure.  
An attempt to consider the effect of wall-wall interactions has been
made using a spring-block model \cite{kovacs}, but this model does not
take correctly into account the long range nature of the dipolar interactions
which underlie the multi-wall effects.

In this article, we consider a system with many rigid parallel 
Bloch domain walls and study their motion, driven by an external 
(triangular) magnetic field, in a
disordered material with strong in-plane uniaxial anisotropy. 
The model is based on the interplay 
between the dipolar and the external field contributions, in the 
presence of structural disorder. Owing to the simplicity of 
the parallel Bloch configuration, the magnetostatic energy can be treated 
analytically. We can calculate perturbatively the  
demagnetizing factor $\kappa$ as a function of the structural 
and geometrical parameters of the system. The 
perturbative calculation perfectly agrees with the results of the simulations
thus representing a link between a macroscopic measurable quantity 
and the microscopic dynamics of the system.
Moreover, dynamic hysteresis is investigated by integrating
numerically the equations of motion of the domain walls. 
We analyze the behavior of the coercive field 
as a function of the applied field frequency, and of the disorder density and 
intensity. 
Furthermore, with our model we can investigate the Barkhausen noise. 
We find that the probability distributions for the size, the duration and the 
amplitude of the Barkhausen avalanches show a power law behavior with a cutoff,
in qualitative agreement with the  available experimental data. These 
results show that multi-domain effects are in principle important, since they give rise to
a non-trivial BN and modify the effective parameters (e.g. the demagnetizing
factor) describing the motion of a single domain wall. A more complete
treatment should involve the dynamics of a system of flexible parallel
domain walls, but this goal is beyond the scope of the present work.

The manuscript is organized as follows. In section \ref{model}
we present an overview on the energetics of films with parallel Bloch
domain walls, and we compute magnetostatic, disorder and external field 
contributions to the equations of motion of the domain walls. In Sec. 
\ref{suscettivita} we present the extended perturbative calculation
of the magnetic susceptivity and the comparison with the simulation results. 
Next, we present the results obtained by our simulations for the analysis of 
the dynamic hysteresis (Sec. \ref{dynamic_hysteresis_p}) and of
the Barkhausen noise (Sec. \ref{barkhausen_noise_p}). 
Finally, our results are resumed in Sec. \ref{conclusioni}.

\section{Interactions in a multi-domain structure}

\label{model}

Our purpose is to study the motion of $n$ parallel Bloch domain walls
in a  disordered ferromagnetic system, under an external magnetic field driving.
The aim is to write an equation of motion for each wall 
and integrate the system of equations numerically. 
To this end, we calculate the total forces acting on the
domain walls whose positions at time $t$
are described by the vector
${\bf x}(t)=\left\{x_0(t),x_1(t),x_2(t),.....,x_n(t)\right\}$, 
As we are interested in the macroscopic 
response, we do not consider the details of the internal structure of the 
walls, and treat only the magnetostatic, the disorder and the external field
contributions. Thus the total force $F(k,t)$ 
acting on the $k$-th wall at time $t$ is given by:

\begin{equation}
F(k,t)=F_{m}(k,t)+F_{dis}(k,t)+F_{ext}(k,t).
\label{Etot}
\end{equation}

In equation (\ref{Etot}), the magnetostatic term $F_{m}(k,t)$ 
takes into account
the interaction between the magnetization and the stray (magnetostatic) field generated by
the magnetic charges due to discontinuity of the magnetization 
vector at the upper and lower boundaries of the sample, $F_{dis}(k,t)$ models
the contribution of structural disorder, impurities, defects and so on, 
and $F_{ext}(k,t)$ describes the interaction between the magnetization and the external
magnetic field. A detailed expression for Eq.~(\ref{Etot}) is obtained by
computing the energy $E=E_m+E_{dis}+E_{ext}$ for a generic
configuration ${\bf x}$ and then deriving it according to:
\begin{equation}
F(k,t)= -\frac{\partial E}{\partial x_k}.
\end{equation}
In the following subsections we will discuss these terms 
in more detail.

\subsection{Magnetostatic force}

\label{magnetostatic_force_calculation}

\begin{figure}[h]

\centerline{\psfig{figure=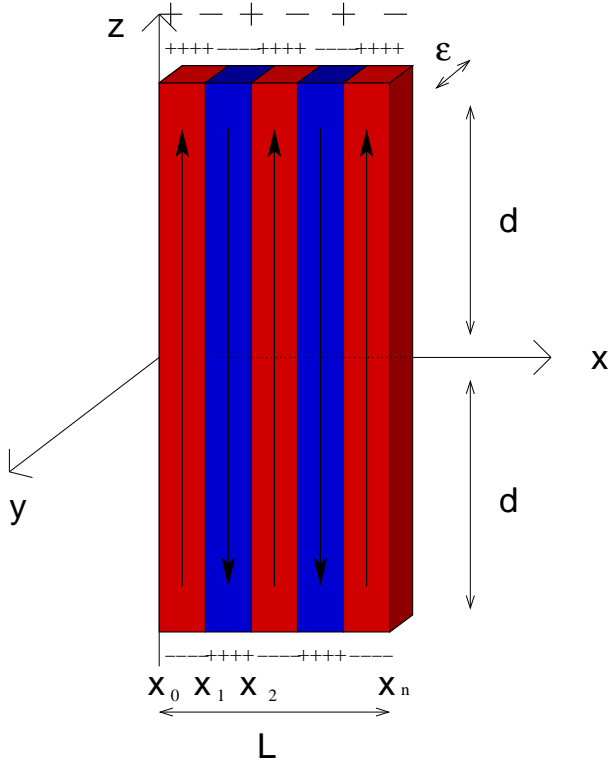,width=8cm,clip=}}

\caption{(Color online) Sketch of the system parameters, for an array of wall positions $\left\{ x_1,x_2,.....,x_n\right\}$.}

\label{sample}

\end{figure}

We consider a sample with total length $L$, height $2d$ and thickness 
$\epsilon$, with $n$ domain walls displaced in the positions 
$\left\{x_0(t),x_1(t),x_2(t),.....,x_n(t)\right\}$ at time $t$, separating
$n-1$ domains of alternating magnetization. 
We consider a system with strong uniaxial in-plane
anisotropy along the $z$ direction. The even domains have a  
magnetization equal to $M_s$ and the odd ones to
$-M_s$, being $M_s$ the saturation magnetization of the material
(see Fig. \ref{sample} for a definition of the parameters).
In order to calculate the magnetostatic contribution $F_m(k,t)$ 
to the total force on the $k$-th domain wall at time $t$,
we first compute the magnetostatic energy of the system $E_m$
for a generic arrangement of the walls and then derive it
with respect to $x_k$.

The magnetostatic energy $E_{m}$ can be expressed in terms of the
demagnetizing field ${\bf H}_d$ as

\begin{equation}
E_{m}=-\frac{1}{2}\,\mu_0\int_{sample}\!\!\!\!\!\!\!\!\!{\bf M}\cdot{\bf H}_d\,\,dxdydz ,
\label{Emagn}
\end{equation}
where ${\bf M}$ is the magnetization vector and the integration is taken over the whole sample.
Eq.~(\ref{Emagn}) can be rewritten as 
$$E_{m}=-\frac{1}{2}\,\mu_0\sum_{i=0}^{n-1}(-1)^i\int_{x_i}^{x_{i+1}}\!\!\!dx\int_0^{\epsilon}dy\int_{-d}^ddz\,M_s\left(\hat{z}\cdot {\bf H}_d\right),
$$
since the magnetization lies on the $z$ axis. 
Using that ${\bf H}_d=-{\overrightarrow \nabla}\phi$, where $\phi$ is the 
scalar potential, it follows that 

$$\hat{z}\cdot{\bf H}_d=-\frac{\partial}{\partial z}\phi(x,y,z) ,$$
and therefore

$$E_{m}=\frac{1}{2}\,\mu_0\sum_{i=0}^{n-1}(-1)^i\int_{x_i}^{x_{i+1}}dx\int_0^{\epsilon}dy\int_{-d}^ddz\, M_s\frac{\partial}{\partial z}\phi(x,y,z)$$

$$=\frac{1}{2}\,\mu_0M_s\sum_{i=0}^{n-1}(-1)^i\int_{x_i}^{x_{i+1}}dx\int_0^{\epsilon}dy\left[\phi(x,y,d)-\phi(x,y,-d)\right]
.$$

The scalar potential is given by the surface integral

$$\displaystyle\phi(x,y,z)=\int\!dS\,\,\frac{\sigma}{r}$$

$$=\sum_{j=0}^{n-1}(-1)^j\int_{x_j}^{x_{j+1}}\!\!\!dx'\int_0^{\epsilon}\!dy'M_s$$

$$\times\left[\frac{1}{\sqrt{(x-x')^2+(y-y')^2+(z-d)^2}}\right. $$

$$\left. -\frac{1}{\sqrt{(x-x')^2+(y-y')^2+(z+d)^2}}\right] ,$$
since the surface charge density $\sigma$ value is $M_s$ or $-M_s$ on the 
upper and lower interfaces $(x,y,\pm d)$ and $0$ elsewhere. 
The magnetization is uniform inside the domains, thus no volume density of 
charge is taken into account.

Thus we obtain

$$E_m=\mu_0 M_s^2\sum_{i,j=0}^{n-1}(-1)^{i+j}\int_{x_i}^{x_{i+1}}\!\!\!dx\int_{x_j}^{x_{j+1}}\!\!\!dx'\int_0^{\epsilon}\!\!\!dy\int_0^{\epsilon}\!\!\!dy'$$

\begin{equation}
\left[\frac{1}{\sqrt{(x-x')^2+(y-y')^2}}-\frac{1}{\sqrt{(x-x')^2+(y-y')^2+4d^2}}\right].
\label{trascuro}
\end{equation}

Assuming that $\epsilon << 2d$, as in most of the samples of experimental interest, 
we can neglect the term $(y-y')^2$ with respect to $4d^2$ in the Eq.~\ref{trascuro}. 
Therefore, the result of the integral is given by

$$E_m=\mu_0 M_s^2\sum_{i,j=0}^{n-1}(-1)^{i+j}\frac{1}{2}$$

$$\left\{\left[g(x_{i+1},x_{j+1})+g(x_i,x_j)-g(x_{i+1},x_j)-g(x_i,x_{j+1})\right]\right.
$$

\begin{equation}
\left.-\epsilon^2\left[f(x_{i+1},x_{j+1})+f(x_i,x_j)-f(x_{i+1},x_j)-f(x_i,x_{j+1})\right]\right\},
\label{eemme}
\end{equation}
where

$$
\left\{
\begin{array}{lc}
\displaystyle
f(x,y)=f(y,x)=2\sqrt{4d^2+(x-y)^2}\\+(x-y)\ln\left(\frac{y-x+\sqrt{4d^2+(x-y)^2}}{x-y+\sqrt{4d^2+(x-y)^2}}\right)\\\\
g(x,y)=g(y,x)=(x-y)\epsilon^2\ln\left(\frac{y-x+\sqrt{\epsilon^2+(x-y)^2}}{x-y+\sqrt{\epsilon^2+(x-y)^2}}\right)\\-\frac{2}{3}|x-y|^3+(x-y)^2\epsilon\ln\left(\frac{\sqrt{(x-y)^2+\epsilon^2}-\epsilon}{\sqrt{(x-y)^2+\epsilon^2}+\epsilon}\right)\\+\frac{2}{3}\left((x-y)^2+\epsilon^2\right)^{3/2}
\end{array}
\right.
$$

\begin{figure}[h]
\centerline{\psfig{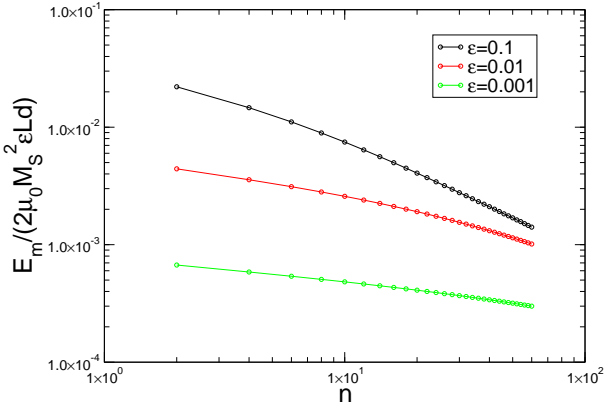}}
\caption{(Color online) Magnetostatic energy $E_m$ as a function of the number of domains
  $n$, for three different values of the system thickness $\epsilon=0.1;0.01;0.001$,
  and for $L=1.$, and $d=10$, calculated for even numbers of domains. Solid lines are guides to the eye.}
\label{E_vs_n}
\end{figure}

To give an idea of the behavior of the magnetostatic energy, in 
Fig. (\ref{E_vs_n}) we have plotted $E_m(t)$ for a sample of fixed length $L$ and $d$ and for three values of the thickness, $\epsilon=0.1;0.01;0.001$, 
as a function of the number of (periodic) domains. As it can be seen, from
a purely magnetostatic point of view, the energy is a decreasing 
function of the number of domains. In fact, the total number of
domains in a sample is determined by the interplay of the magnetostatic 
term and the domain wall energy, linearly increasing as a function of $n$
~\cite{bertotti}.

The magnetostatic force on the $k$-th wall at time $t$ is thus

$$F_m(k,t)=-\frac{\partial E_m}{\partial x_k(t)}$$

$$=2\mu_0M_s^2\sum_{i=0}^{n-1}(-1)^{i+k}\left[\frac{\partial g(x_{i+1},x_k)}{\partial x_k}-\frac{\partial g(x_i,x_k)}{\partial x_k}\right.$$

\begin{equation}
\left.-\epsilon^2\left(\frac{\partial f(x_{i+1},x_k)}{\partial x_k}-\frac{\partial f(x_i,x_k)}{\partial x_k}\right)\right],
\label{Fmk}
\end{equation}
where

$$\frac{\partial f(x,y)}{\partial y}=-\frac{\partial f(x,y)}{\partial x}=\ln\left(\frac{y-x+\sqrt{4d^2+(x-y)^2}}{x-y+\sqrt{4d^2+(x-y)^2}}\right)$$

$$\frac{\partial g(x,y)}{\partial y}=-\frac{\partial g(x,y)}{\partial x}=2(x-y)\sqrt{\epsilon^2+(x-y)^2}$$

$$+2\epsilon(x-y)\ln\left(\frac{\sqrt{\epsilon^2+(x-y)^2}-\epsilon}{\sqrt{\epsilon^2+(x-y)^2}+\epsilon}\right)$$

\begin{equation}
+\epsilon^2\ln\left(\frac{y-x+\sqrt{\epsilon^2+(x-y)^2}}{x-y+\sqrt{\epsilon^2+(x-y)^2}}\right)-2(x-y)^2sign(x-y).
\label{derivate}
\end{equation}

\begin{figure}[h]
\centerline{\psfig{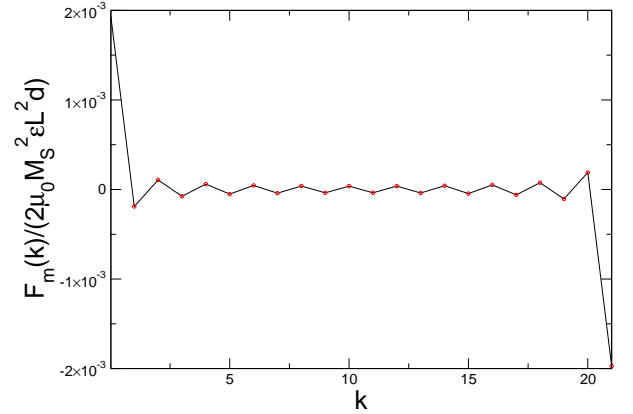}} 
\caption{(Color online) Magnetostatic force $F_m(k,t)$ on the $k$-th domain wall as a
  function of $k$ (red dots), for a periodic configuration of the domain walls
  positions. The sample has $n=21$ domains, $\epsilon=0.002$, $d=10$, and
  $L=1$. The solid line is a guide to the eye.}
\label{forza}
\end{figure}

In Fig. (\ref{forza}) we show the magnetostatic force  $F_m(k,t)$ calculated 
at each wall $k$ for a periodic configuration of $n=21$ domains. 
As it could be seen, walls with $k$ of opposite parities are driven in opposite
directions, and the absolute value of  $F_m(k,t)$ is higher for the external
walls, which is a finite size effect tending to reduce the size of the 
boundary domains. For a large odd number of domains, in the center of the sample,
all the $F_m(k,t)$ would be equal in magnitude, as it can be seen 
in Fig. (\ref{forza}). For an even number of domains, the central wall experiences a zero force for 
symmetry reasons. As expected, for a generic (even or odd) number of
domains, the effect of the magnetostatic force is to move the walls in the 
direction that decreases the 
magnetization, in order to minimize the magnetostatic energy 
[see Eq. (\ref{Emagn})].

\subsection{Disorder}

Different sources of inhomogeneities are found in virtually all ferromagnetic
materials. The presence of structural disorder is essential to 
understand the hysteretic behavior, and especially to account for the residual
coercive field when the frequency of the external driving field vanishes. 
Disorder is provided by vacancies, non-magnetic impurities, dislocations or grain
boundaries in crystalline systems, variations of the easy axis between
different grains for polycrystalline materials, and internal stresses
for amorphous alloys.

We consider here only quenched (frozen)
disorder that does not evolve on the timescale of the magnetization
reversal, which is usually a realistic approximation for ferromagnetic systems. At the beginning of a simulation, we extract the positions
of a fixed number $n_p$ of pinning centers with a uniform probability 
distribution all over the sample. Each pinning center interacts
with every domain wall by a located potential of the type

\begin{equation}
U_p(x)= A \exp[-(x/\xi)^2],
\end{equation}
where $x$ is the distance between the pinning center and the
wall, $\xi$ is a correlation length and the amplitude $A$ is extracted from a uniform distribution, 
considering that the strength of the pinning centers should
vary for structural reasons. In this way we produce a disordered
landscape in which the domain walls evolve.

\subsection{External field}

The interaction between a ferromagnetic system
and an external magnetic field ${\bf H}$ is described by the energy term:

\begin{equation}
E_{ext}=-\mu_0 \int_{sample}{\bf M}\cdot {\bf H}\,dV .
\label{intext}
\end{equation}

In our model the uniaxial anisotropy and 
the external time dependent magnetic field $H_{ext}(t)$ are parallel to the 
easy axis of the sample, and thus parallel or anti-parallel 
to the magnetization. 
Thus in this case the total energy is
$$E_{ext}=2d\epsilon \mu_0M_sH_{ext}\sum_{i=0}^{n-1}(-1)^i\left( x_{i+1}(t)-x_i(t)\right),$$
and the external field contribution $F_{ext}(k,t)$ to the force $F(k,t)$ 
on the $k$-th wall at time $t$ is given by the partial 
derivative of Eq. (\ref{intext}) with respect to the 
wall position $x_k(t)$, and could be thus written as

\begin{equation}
F_{ext}(k,t)=4\mu_0 M_s d \epsilon H_{ext}(t)(-1)^{k+1},
\label{intext_parallele}
\end{equation}
which depends of course on the parity of the domain wall.

\section{Demagnetizing factor and magnetostatic susceptivity}

\label{suscettivita}

An interesting analytical result that could be obtained from our model
is the dependence of the demagnetizing factor $\kappa$ on the number
$n$ of the parallel domain walls. 
The  parameter $\kappa$ enters as a mean field term
in several domain wall models used to describe the BN 
\cite{ABBM,zapperi,zapperi2,urbach,bahiana}. Its precise dependence
on the sample geometry is usually approximated considering a uniformly
magnetized sample and the domain structure is ignored.  
Understanding
this point is important since $\kappa$ determines for instance 
the size dependence
of the Barkhausen avalanche distribution \cite{dz}.

The demagnetizing factor $\kappa$ can be rigorously defined 
for uniformly magnetized ellipsoid from the relation
$H_d=-\kappa M$. In more general situations, we can define it as an
effective quantity linking the average demagnetizing field
to the magnetization $\kappa \equiv -\langle H_d \rangle/M$.
Here we compute $\kappa$ from the magnetic response of the system 
to a small perturbative external field. Since at equilibrium
a small change in the external field is compensated by 
an increase in the demagnetizing field ($H_{d}+H_{ext}=0$), we can link
$\kappa$ to the magnetostatic susceptivity $\chi=dM/dH_{ext}$
by the relation $\kappa=1/\chi$.

For an extended system with a large and even number of domains $n$,
the equilibrium configuration consists of a domain wall array
$$x_i=\frac{iL}{n}\,\,\,\,\,\,\,\,\, \forall i$$
(note: in what follows we will omit the $t$ dependence of the functions).
We can choose a positive perturbative field $\delta H_{ext}$ and 
thus consider the perturbed walls positions:

$$
\left\{
\begin{array}{l}
\displaystyle
x_i=\frac{iL}{n}+(-1)^{i+1}u\\\\
x_0=0\\\\
x_n=L
\end{array}
\right.
$$
that correspond to a (small) increase of the magnetization $dM\sim n u$, with
$u>0$, 
uniformly distributed over the domain walls.
Since the calculation of the magnetic susceptivity is quite involved, we
discuss it in the Appendix and report here only the final result: 
\begin{equation}
\chi=\frac{\partial M}{\partial H_{ext}}=\frac{1}{\kappa}=\frac{d}{\epsilon A(n)},
\label{chi}
\end{equation}
where $$A(n)=2\left[\gamma+\ln(n)\right],$$
and $\gamma\simeq 0.577215$. Higher order corrections in $n$ have been neglected.

\begin{figure}[h]
\centerline{\psfig{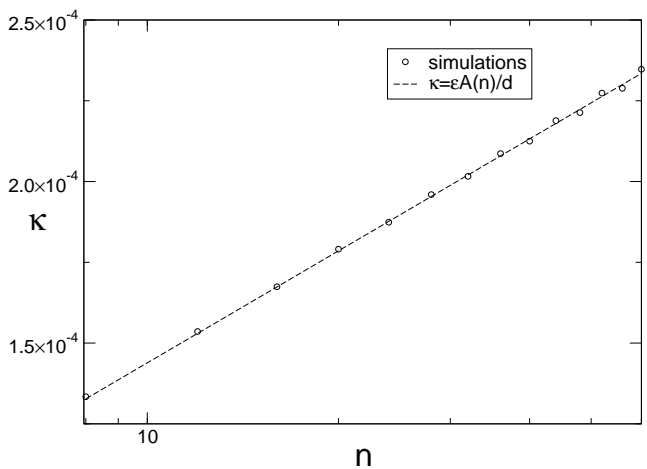}} 
\caption{Demagnetizing factor $\kappa$ as a function of the number of  domain
  walls $n$ (semilogarithmic scale):
  comparison between simulation results and perturbative calculation (dashed line) for
  $d=400$, $\epsilon=0.01$, and for unitary $L$.}
\label{confronto}
\end{figure}

In Fig. \ref{confronto} we compare the results of our simulations
and the theoretical result of Eq. \ref{chi}, for a system with $d=400$,
$\epsilon=0.01$ and $L=1$, without disorder, 
by studying $\kappa$ as a function of the number of domain walls $n$ 
(coinciding with the number of domains in the limit $n\rightarrow \infty$). The simulations are performed
starting from a periodic configuration of a high and even number of domains,
and letting the walls relax under an external field that is very low, 
in order to have a displacement from the equilibrium position 
of the walls of the order
of some per cents of the distance between nearest-neighbor walls.
To fulfill the hypothesis under which Eq. \ref{chi} is calculated, 
we use here periodic boundary conditions.
As it can be seen, the agreement between theory and simulations 
is excellent, even if the simulations have been performed for 
finite samples. 
The value of the magnetic susceptivity is of the order $1/\kappa=\chi\simeq10^4$,
which is a realistic value for a ferromagnetic system.
It would be very interesting to test the prediction of  Eq. \ref{chi} with
experimental results on materials with low disorder.
Finally, it is interesting to notice that the demagnetizing factor
$\kappa=1/\chi$ displays an inverse dependence on the sample length $d$
and a linear dependence on the thickness $\epsilon$. Hence, as expected,
reducing the sample aspect ratio leads to an increase in $\kappa$. This
fact has been exploited in Ref.~\onlinecite{dz} to tune $\kappa$
and study its effect on the BN.

\section{Numerical model}
\label{simulations_p}

In the previous sections we have calculated the various contributions to 
the total force acting on each wall (Eq. \ref{Etot})
thus it is now possible to simulate the dynamics of the system. 
Since the number of the walls is  $n$, and we consider in this part closed boundary conditions (the boundary walls do not move),  we have to write down $n-2$ equations of motion given by
\begin{equation}
\Gamma \frac{dx_k}{dt}=F(k,t)=F_m(k,t)+F_{ext}(k,t)+F_{dis}(k,t).
\label{equazione_moto_p}
\end{equation}
In the following we set the damping coefficient $\Gamma$ to unity. 
We integrate these equations using a fourth order Runge-Kutta algorithm. 
The external magnetic field is a saw-tooth signal with rate $\omega$,
being $\dot{H}\propto\omega$.

We always start our 
simulations with an $M=0$ at $H_{ext}=0$ periodical configuration, and
drive the sample to positive and then to negative saturation.
We include the possibility of nucleation and annihilation of the domain walls
by fixing a minimum interaction range between two nearest-neighbor 
walls, namely $\delta_{min}$. If two walls get closer than 
$\delta_{min}$, we stop them and consider the pair of walls as annihilated. 
This means that their contribution on the sum involved in the 
calculation of long-range magnetostatic force (see Eq. \ref{Fmk}), 
and thus their effect on the equations of motion of the other walls, 
is no more considered. Besides, we go on calculating the total force on the 
annihilated pair of walls. When this force become strong enough (and of the
correct sign) to let them escape from the range $\delta_{min}$, we nucleate
them and consider again their contribution to the magnetostatic force
calculation, Eq. \ref{Fmk}.
This corresponds to have a nucleation barrier equal to zero.
Since we are interested in the study of Barkhausen noise and of the 
coercive field behavior connected to dynamic hysteresis, 
this choice does not influence our results,
as we have checked.
In fact, all the phenomena that we are interested in take place in the 
central region of the hysteresis loop, namely the one close to 
the coercive field $H_c$, while the choice of the nucleation barrier
involves only the region of the cycle close to  saturation.
Moreover, for the same reason, our results are  independent 
on the choice of $\delta_{min}$, as long as it is chosen in a reasonable range.

\section{Coercive field and dynamic hysteresis}
\label{dynamic_hysteresis_p}
\subsection{Dynamic hysteresis}

\begin{figure}[h]
\begin{center}
\begin{tabular}{c}
\includegraphics[height=6.5cm]{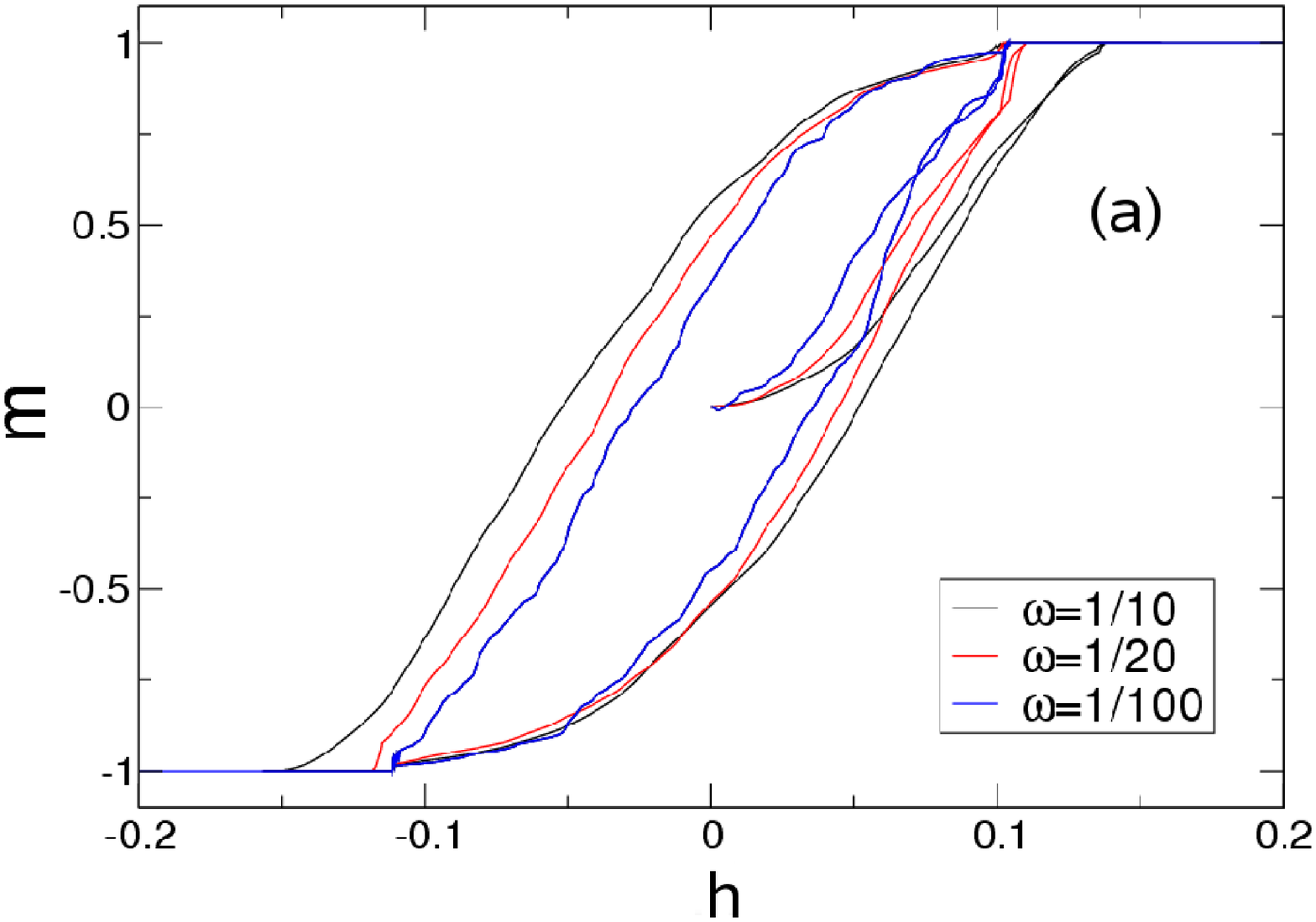} \\
\includegraphics[height=5.5cm]{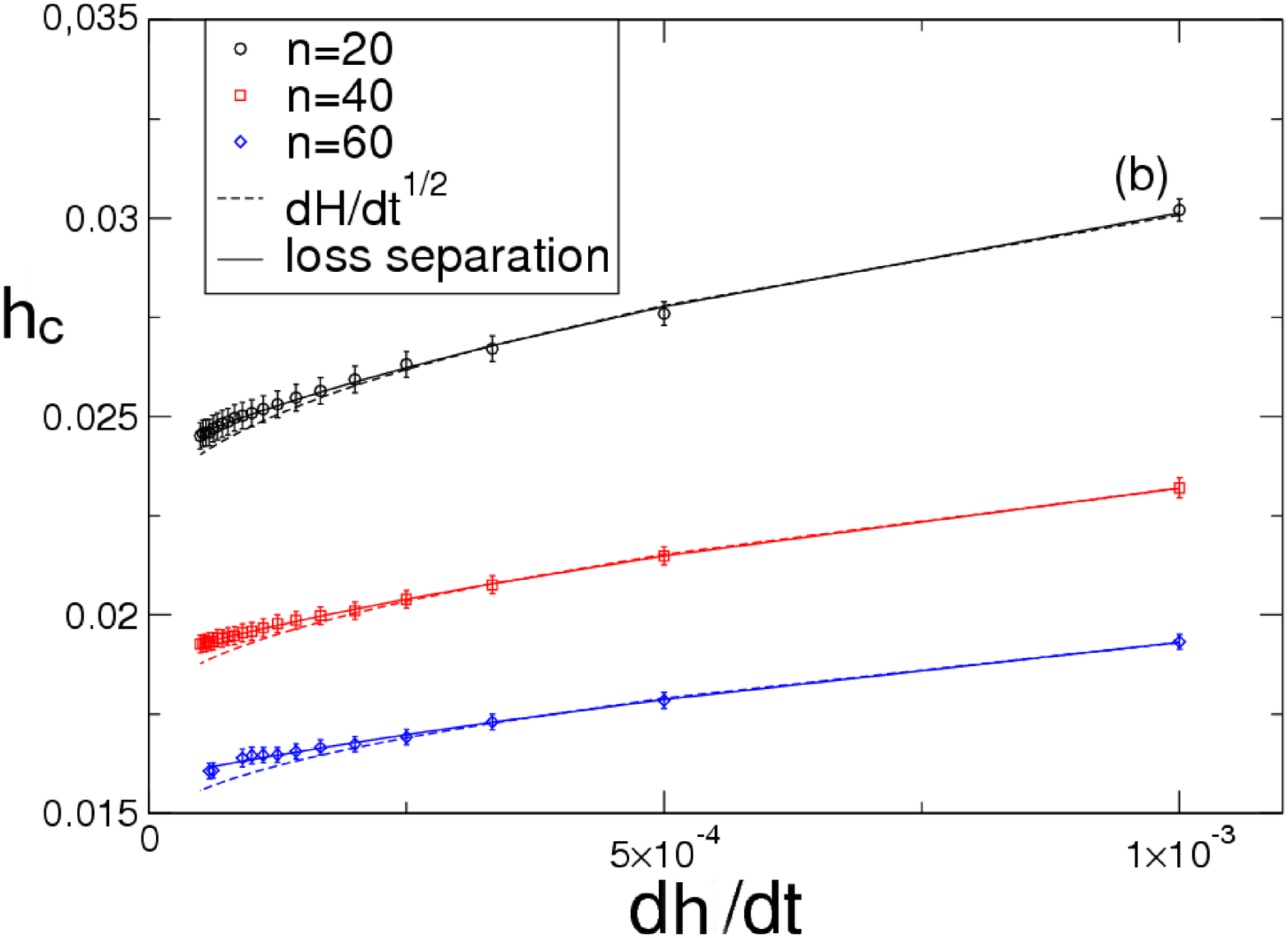} 
\end{tabular}
\end{center}
\caption{(Color online)
(a) Hysteresis loops for different external field rates, for a system
with $n=20$ domains, and (b)
Coercive field as a function of the external field rate
  for three different system sizes, $n=20,40,60$; $h_c$ vs
  $dh/dt$ in a linear plot, in comparison with a
  square root fit (dashed line) and the fit with the power loss formula
  (continuous line, see text). Every point is mediated over $100$ realizations
  corresponding to different disorder configurations.}
\label{Hc_vs_frequenza_p}
\end{figure}

We first consider the effect of the external field rate on 
the hysteretic behavior. In Fig. \ref{Hc_vs_frequenza_p}(a) 
we show the hysteresis loops obtained for various rates 
for a system with $n=20$ domains (since from now we will deal with a
reasonably high number of domains, we will use with a slight abuse of terms 
$n$ for the number of domains), where $m$ is the normalized magnetization
$m=M/M_s$ and $h$ is the normalized magnetic field $h=H/M_s$. As expected from experiments 
and general considerations, small (high) frequencies correspond to narrow
(large) cycles. To quantify this observation we can focus on the coercive field
$H_c$ behavior. In Fig. \ref{Hc_vs_frequenza_p}(b) we show the dependence 
of $H_c$ on the field rate $\omega$.

For all the considered values of $n$
in Fig. \ref{Hc_vs_frequenza_p}, $n=20,40,60$ (for a
discussion of the behavior of $H_c$ as a function of the number of domains
$n$ see the next subsection), $H_c$ shows an increasing dependence
on $\omega$. The $H_c$ dependence on $\omega$ is quite close to a power law
 of the form 

\begin{equation}
H_c=H_p+A\omega^{1/2},
\label{sqrtomega_p}
\end{equation}
as suggested for experiments on permalloy thin films and microstructures fabricated from the same
sample \cite{nistor}, 
and by the theory presented in Ref.~\onlinecite{lyuksyutov}.

Nevertheless, as it can be seen in Fig.\ref{Hc_vs_frequenza_p}(b), our curves are best fitted by the loss separation formula 
\cite{bertotti,colaiori},

\begin{equation}
H_c=H_p+C_{ex}[(1+r\dot{H})^{1/2}-1],
\label{lossse}
\end{equation}
where $C_{ex}=n_0V_0/2$, and $r=4\Gamma\mu/(n_0^2V_0)$. Here $n_0$ is the number of
active walls in the  quasi-static limit ($\omega\rightarrow 0$), $V_0$ is a characteristic field which controls the
increase of $n_0$
due to the excess field, $\mu$ is the permeability
and $H_p$ is the static (hysteretic) component.
The fit parameters values are shown in table \ref{tabella}.

\begin{table}[h]
 \begin{center}
\begin{tabular}{|c|c|c|c|}
\hline
n&20&40&60\\ 
\hline
$H_p$&0.024&0.019&0.016 \\
$C_{ex}$&0.003&0.003&0.0036 \\
r&6622&4752&2989\\
$\mu$&183&153&173\\
$n_0$&18&21&32\\
$V_0$&0.0003&0.0003&0.0002\\
\hline
\end{tabular}
\caption{\label{tabella} Fit parameter values for Fig. \ref{Hc_vs_frequenza_p}(b).}
\end{center}
\end{table}

The behavior of $H_c$ as a function of 
$\omega$, suggested in Eq. \ref{lossse},
means that in the adiabatic limit (low frequencies) $H_c$ goes to a 
non-vanishing value $H_p$ 
that we can interpret as the pinning dominated quasi-static component
due to structural disorder, while the behavior of $H_c$ in the high
rate regime represents the domain wall dominated dynamic contribution.
Moreover, the scaling function of Eq. \ref{lossse} 
describes the evolution from the 
adiabatic to the domain wall dominated power-law behavior without invoking
mechanism crossover between domain wall propagation and nucleation of new
domains. In fact, even if the nucleation process is included in our model,
besides in a very simplified form, 
nucleation takes place only in the regions of the hysteresis loops very 
close to saturation. Thus the cycles are very oblique [see Fig. 
\ref{Hc_vs_frequenza_p}(a)]. Such a kind of hysteresis loops are expected to be dominated by domain wall motion.

\subsection{Role of the number of domains}

\begin{figure}[h]
\begin{center}
\begin{tabular}{c}
\includegraphics[width=9.cm]{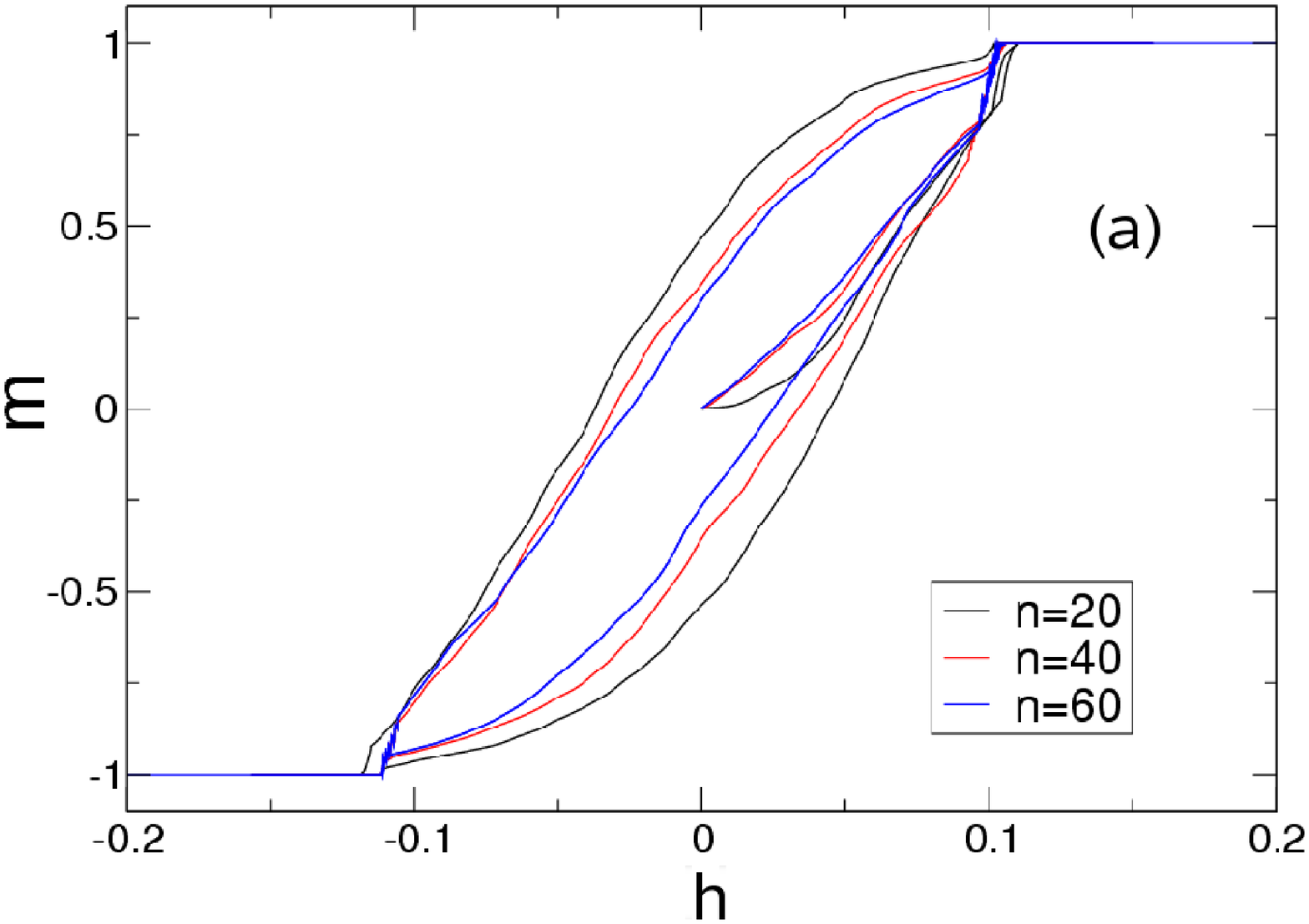}\\

\includegraphics[width=9.cm]{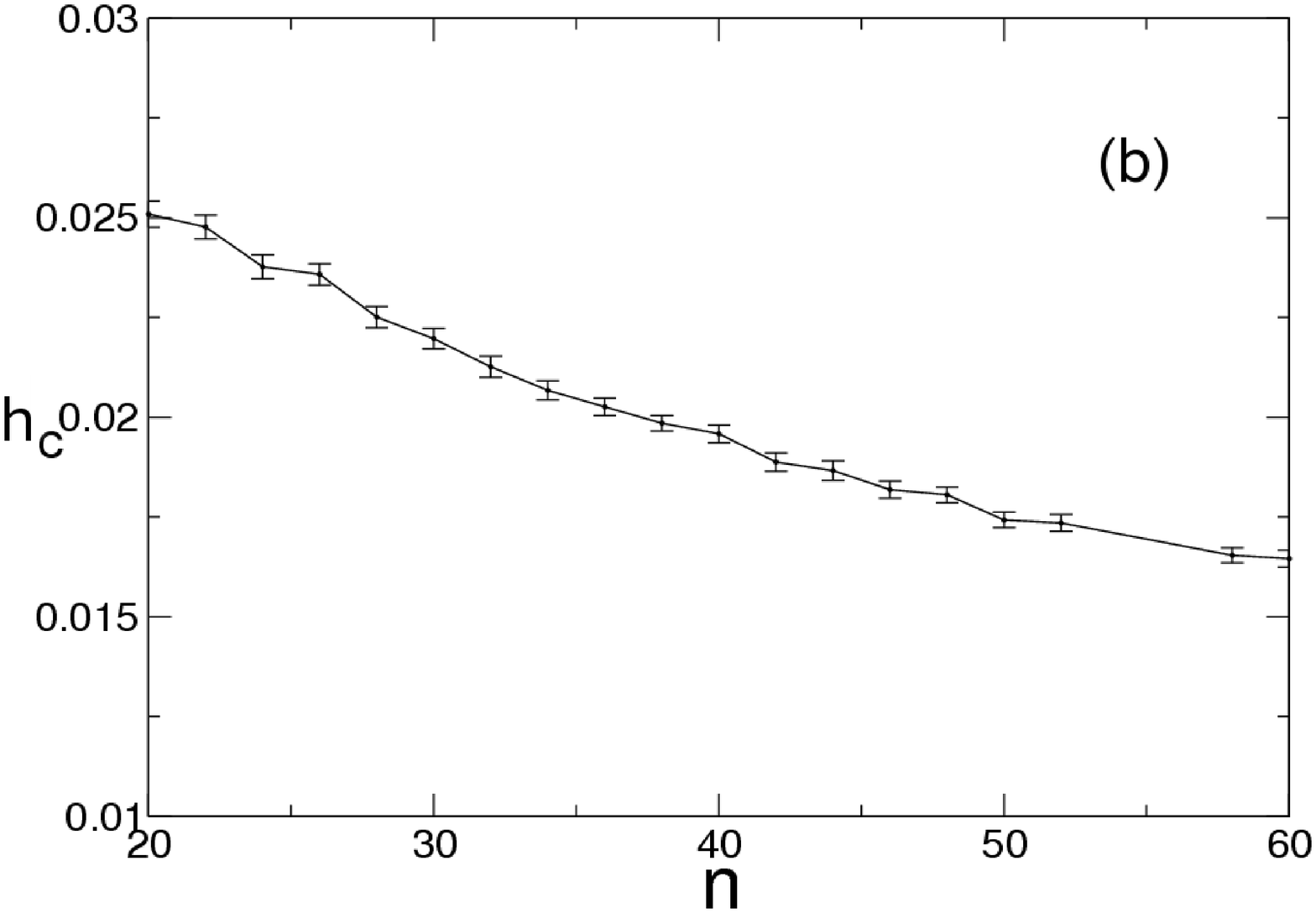} 
\end{tabular}
\end{center}
\caption{(Color online) (a) Some hysteresis loops for different system sizes, $n=20,40,60$, for an applied field rate $\omega=1/20$, and (b)
  Coercive field $H_c$ as a function of the number of domains $n$. Every point
  is an average on $100$ samples, corresponding to different disorder
  realization, for an applied field rate $\omega=1/1000$. The line is a guide
  to the eye.}
\label{nvar}
\end{figure}

Another interesting issue to analyze is the effect of the number of 
domains $n$ on the system behavior. We remind here that at the beginning of each
simulation we set the maximum number of domains of the specimen. 
This number can decrease down to one while driving the system to 
saturation, and then, increasing the external field, it is 
completely restored around the coercive field. From a qualitative point of view, 
keeping constant the other system parameters (i.e. the physical
dimensions of the specimen under study and the pinning centers density and strength $f_0$), 
a high (low) number of domains corresponds to hysteresis loops with a lower
(higher) permeability, in accord with Eq.\ref{chi},
as it can be seen in Fig. \ref{nvar}(a). This happens because the increase of the domain number increases the
relative relevance, in the total force balance, of the dipolar contribution.
In fact, an absence of collective behavior leads to square shaped cycles,
where the whole magnetization reversal process takes place in a few
avalanches. Otherwise, in materials in which dipolar interactions are relevant,
oblique hysteresis loops are expected. In our simulations, the coercive field 
$H_c$ decreases as a function of $n$, as expected [see Fig. \ref{nvar}(b)].

\subsection{Effect of disorder}

As we have already mentioned, the presence of structural disorder has a 
crucial role in the physics of ferromagnetic systems. Disorder 
pins the domain walls, that thus are not able to move 
until the external field reaches a
sufficient value to overcome the pinning and let the wall move again. 
Hence we will expect, from this simple argument, that more
disordered systems have larger hysteresis loops compared to purer
specimens, for the same $\omega$.

\begin{figure}[h]
\begin{center}
\begin{tabular}{c}
\includegraphics[height=8cm,angle=-90]{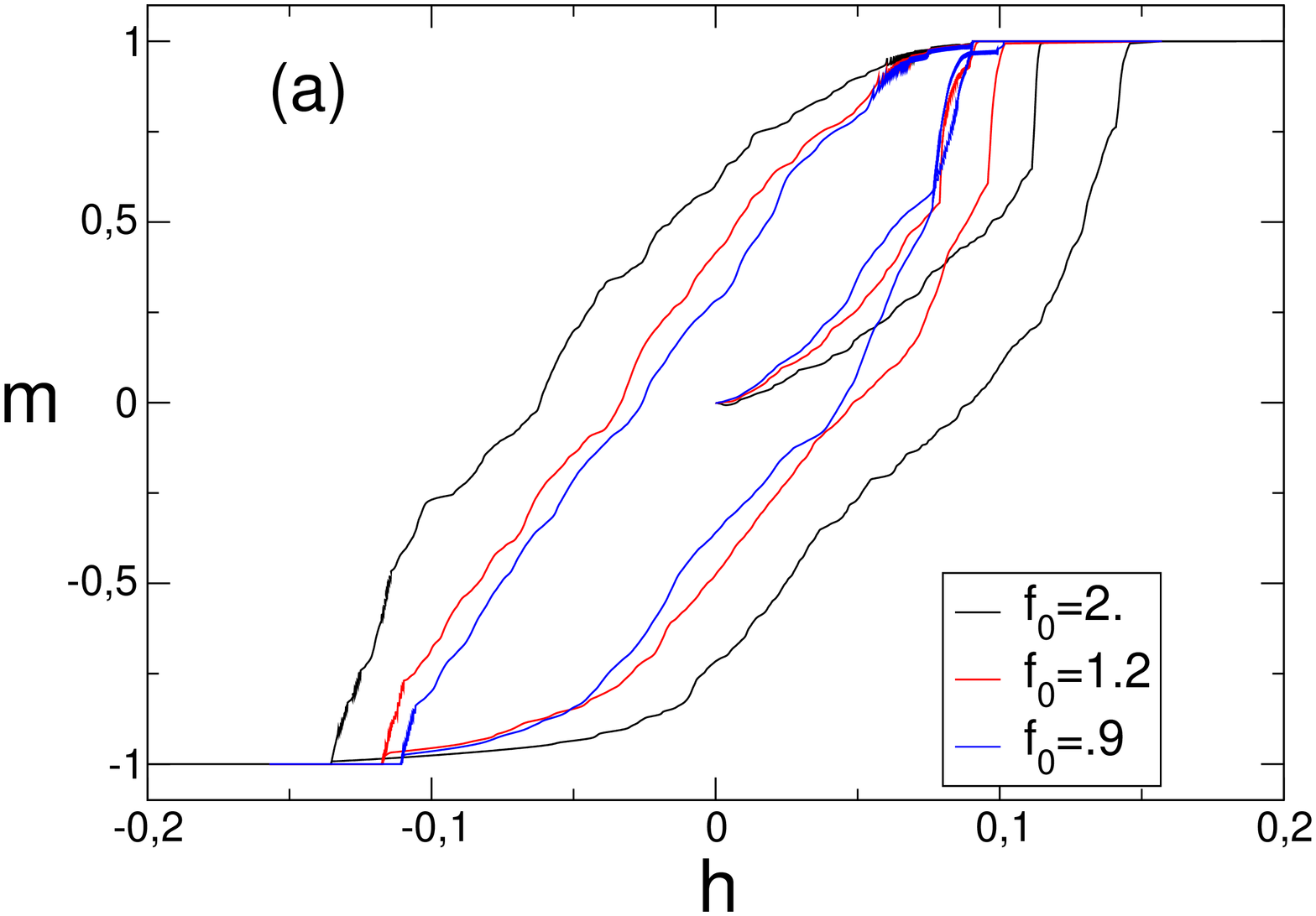}\\
\includegraphics[height=8cm,angle=-90]{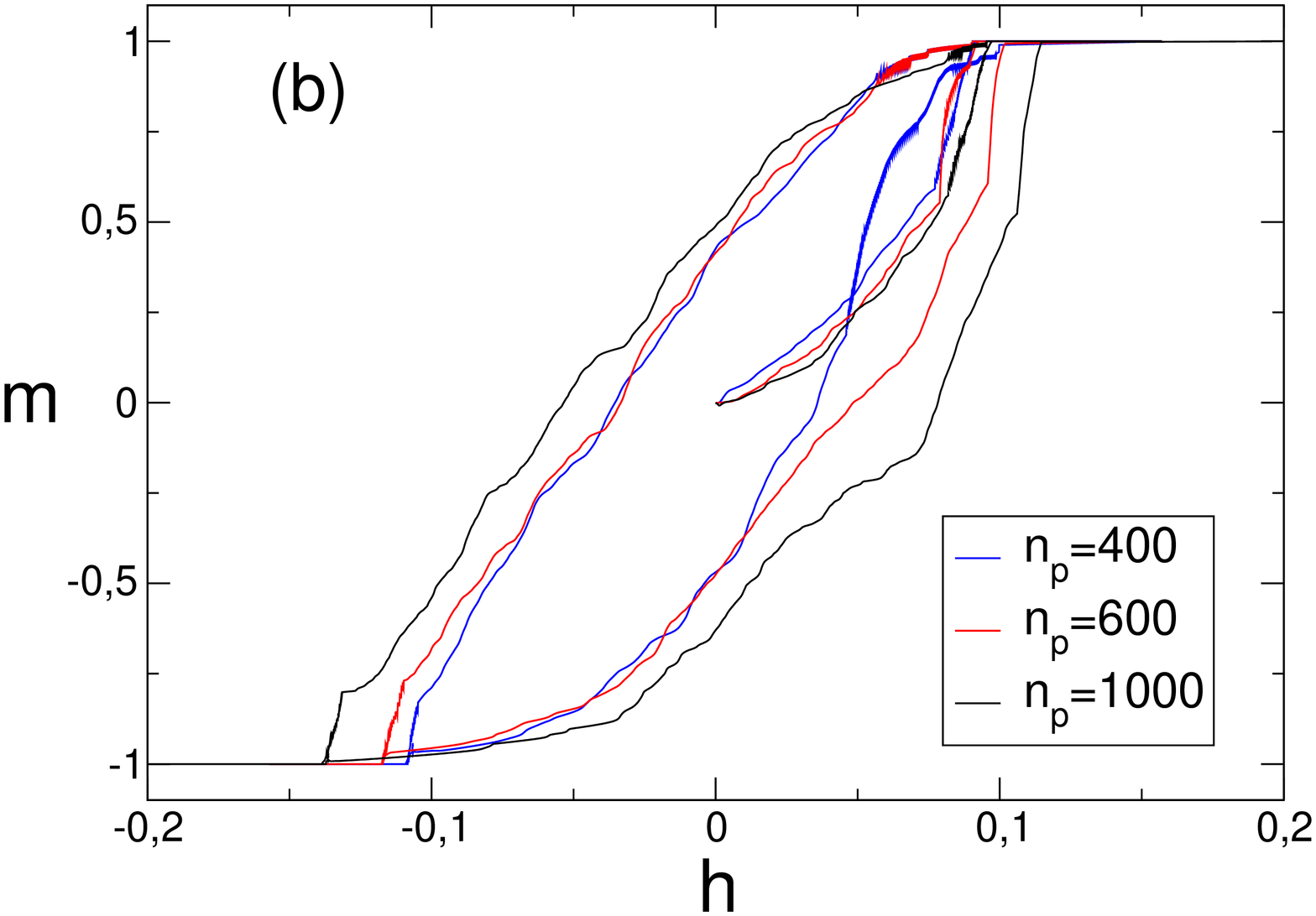} 
\end{tabular}
\end{center}
\caption{(Color online) hysteresis loops for varying: (a) disorder strength $f_0$ ($n_p=600$) and (b) number of pinning centers $n_p$ ($f_0=1.2$). The rate is $\omega=1/50$.}
\label{dis_var_p}
\end{figure}

We remind here that it is difficult to quantify the disorder contribution in real systems.
In our model we have two degrees of freedom to control the disorder:
we can tune the density of pinning centers (by controlling the total number of pinning
sites $n_p$) and their strength $f_0$.

In Fig. \ref{dis_var_p} we show some hysteresis 
cycles for varying values of the two parameters $f_0$ [see panel (a)] and $n_p$ [see panel (b)].
 As we can see, their effect on the 
loops is qualitatively similar, and the behavior for increasing disorder (i.e. 
larger cycles for larger disorder) is in agreement with the general argument explained above.

\begin{figure}[h]
\includegraphics[height=9.cm,angle=-90]{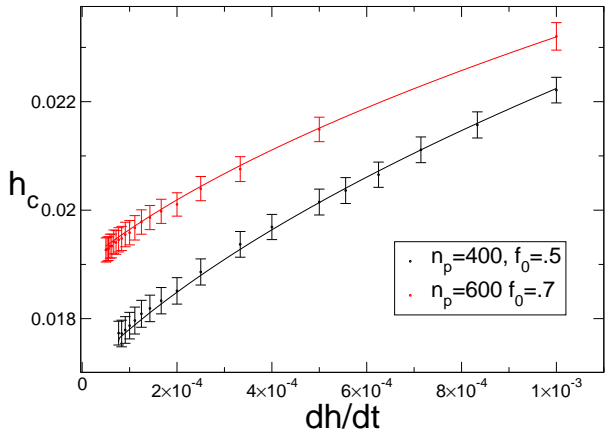}
\caption{(Color online) The reduced coercive field $h_c$ dependence on the
  external field rate $dh/dt$ for a system with $n=40$ domains and two
  different number of pinning centers $n_p$ and disorder amplitude $f_0$. The
  applied field rate is $\omega=1/1000$. Data are fitted by
  Eq.\ref{lossse}. Fit parameters are $H_p=0.019$, $C_{ex}=0.003$ and $r=4752$
  for $n_p=400$; $H_p=0.017$, $C_{ex}=0.0038$ and $r=4752$ for $n_p=600$.}
\label{2dis}
\end{figure}

In Fig. \ref{2dis}, we show the coercive field behavior for increasing disorder, intended here as increasing both the values of $n_p$ and $f_0$: 
as expected, stronger disorder implies
higher coercive field. Moreover, it is important to notice that even varying the disorder strength or density, the $H_c$ dependence on the square root of the 
applied field rate does not change.

\section{Barkhausen noise}
\label{barkhausen_noise_p}

The Barkhausen noise is an unavoidable feature of magnetic hysteresis in
disordered samples and its statistical properties have been widely
studied in experiments and models \cite{dz}. The experiments in
soft ferromagnetic materials have been divided into universality
classes depending on the avalanche distributions \cite{dz}.  
Experimental results are well described by models involving a single flexible
domain wall, and ignoring any interactions between different walls. Different
universality classes are set by the dominant interactions between parts of the
walls. Here we explicitely ignore these local interactions and want to account
only for the interactions between walls.

\begin{figure}[h]
\includegraphics[height=8cm]{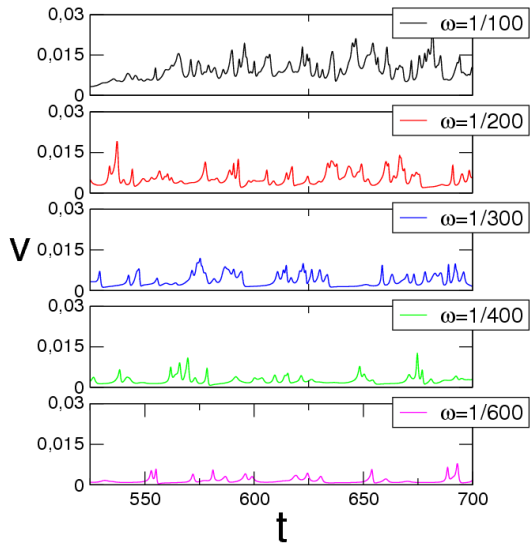}
\caption{(Color online) Time sequences of Barkhausen noise. The labels indicate the applied field rate. We report portions of simulated signals corresponding to the same time interval. The time is measured considering an integrating time step $\Delta t=0.05$.}
\label{segnali}
\end{figure}

In Fig. \ref{segnali} we show some simulated signals for various rates
of the external field. Partial regions are chosen for the signals, 
corresponding to the same time interval. As it can be seen, our results 
reproduce a well-known characteristic of the experimental Barkhausen noise:
at low frequencies, the signal is composed of separated avalanches,
while this property is progressively lost at higher frequencies. In the latter case, the 
signal appears as a continuous sequence of peaks and the single avalanches
are no more distinguishable. We have analyzed the Barkhausen signal 
considering
the distribution of the signal amplitude, avalanche sizes and durations.

\begin{figure}[h]
\begin{center}
\begin{tabular}{c}
\includegraphics[height=9cm,angle=-90]{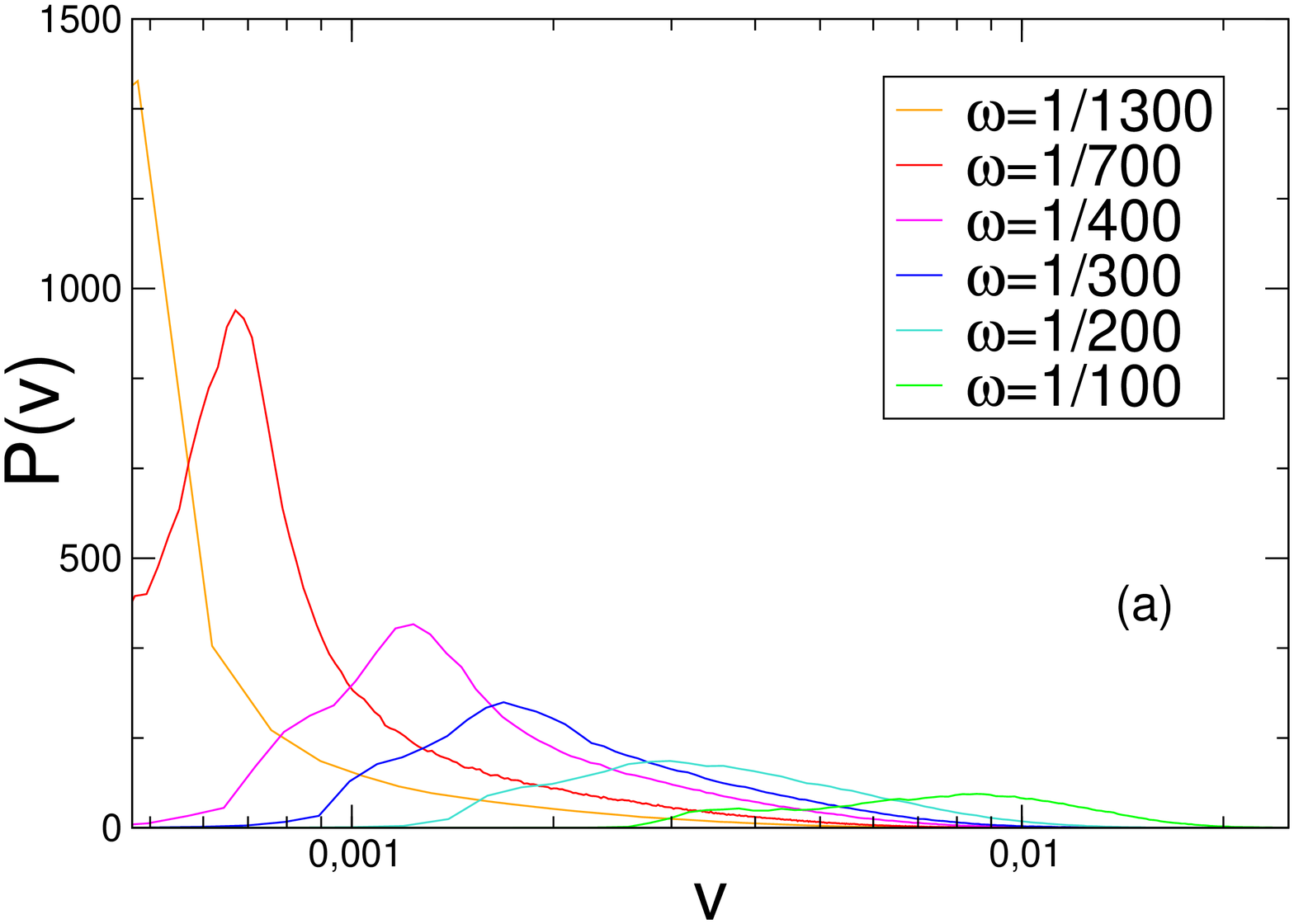}\\
\includegraphics[height=5.5cm]{Fig10b.eps} 
\end{tabular}
\end{center}
\caption{(Color online) (a) Probability distribution of the signal amplitude
  (i.e. of the domain
  walls velocity) $v$, for various frequencies $\omega$ of the external
  magnetic field ($\omega=1/1300,1/700,1/400,1/300,1/200,1/100$, for a system
  with $n=60$ domains. (b) Probability distribution of the signal amplitude for
  $n=60$ and $\omega=1/130$, fitted with the ABBM function,
  $P(v)=av^{c-1}exp(-cv/<\!v\!>)$.}
\label{PdiV_p}
\end{figure}

The amplitude probability distribution of the magnetic induction 
flux rate $\dot{\phi}$ is a measurable quantity and thus has been often
used in order to test the reliability of a model for Barkhausen noise
(see for instance Ref.~\onlinecite{ABBM}). In fact, the flux rate is related to the 
velocity of the domain walls $v$.
In Fig. \ref{PdiV_p}(a) we show the rate dependence of the probability 
amplitude distribution. In the non-adiabatic limit, in fact, the domain
walls velocity is expected to depend on $\omega$. We notice that $P(v)$
passes from a almost symmetric shape for very high frequencies to an 
asymmetric one for low frequencies.

In the ABBM model \cite{ABBM}, the predicted shape of the amplitude probability
distribution is a Gamma function
\begin{equation}
P(v)\propto v^{c-1}e^{-cv/\langle\!v\!\rangle}.
\label{PdiFi}
\end{equation}
We remind here that this formula is obtained under the assumption that the
disorder is a correlated (Brownian) process.

The mechanism giving rise to the Barkhausen effect is the domain walls motion, and that a domain wall motion can be described as a stationary Markov process, associated
with fixed and well-defined values of the magnetization rate imposed to the system
and of the permeability of the specimen $\mu$, associated with the part of the hysteresis
loop where the domain wall motion is assumed to take place.
Even if our model is different from the single wall model ABBM, the formula of Eq.
\ref{PdiFi} is able to fit our probability distributions $P(v)$ for high
rates, although it does not supply a suitable explanation for the 
low rate regime [see Fig. \ref{PdiV_p}(b)].

\begin{figure}[h]
\begin{center}
\begin{tabular}{c}
\includegraphics[height=9cm,angle=-90]{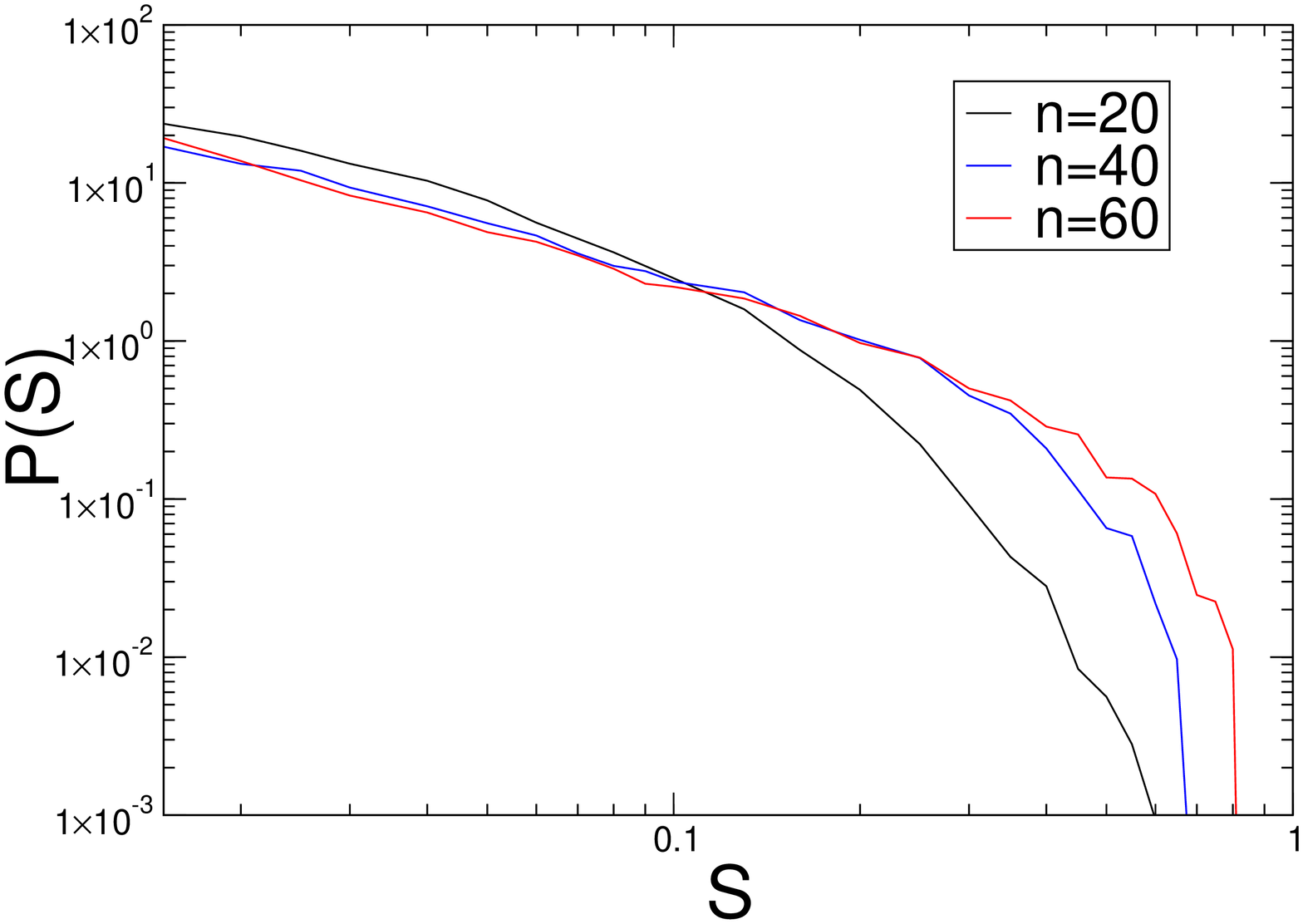}\\
\includegraphics[height=9cm,angle=-90]{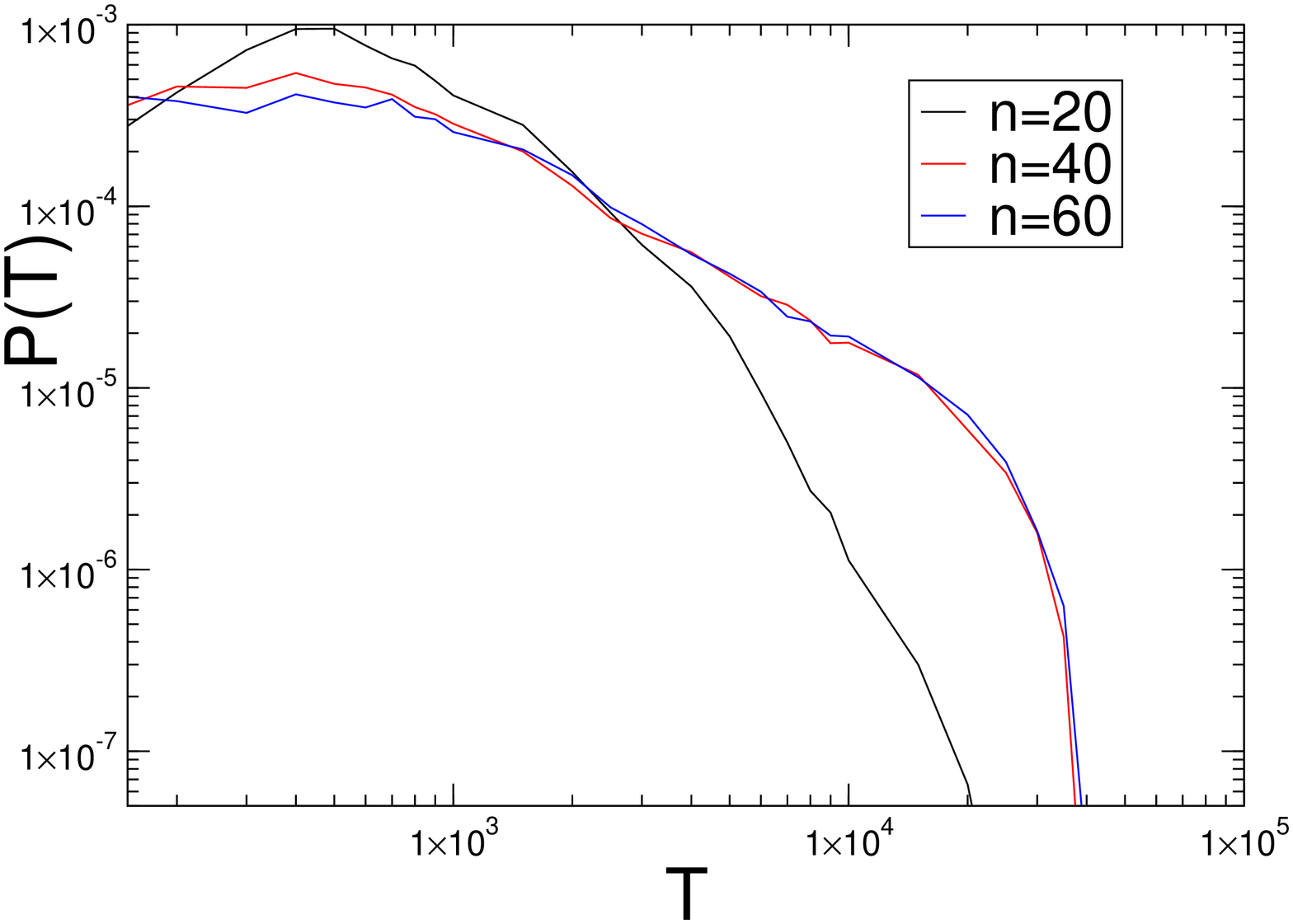} 
\end{tabular}
\end{center}
\caption{(Color online) (a) Probability distribution of the avalanche size $S$, and (b) Probability distribution of the avalanche duration $T$, for three different system sizes ($n=20,40,60$).}
\label{Pdi}
\end{figure}

Next, we evaluate the probability distribution for the duration $T$ and
the size $S\equiv \int^T dt v(t)$ of the avalanches. 
The distributions are shown in Figs. \ref{Pdi}(a) and \ref{Pdi}(b).

As it can be seen, both distributions display a power law behavior
of the type $P(S)\sim S^{-\tau}$ and $P(T)\sim T^{-\alpha}$,
for almost two decades for the maximum studied size ($n=60$), with a cutoff.
The scaling exponents of these distributions are respectively
$\tau\sim 1.1$ and $\alpha\sim 1.2$. These exponent values, however,
have never been reported for bulk soft magnetic materials where parallel
domain walls are commonly observed \cite{dz}. This is not surprising as we
neglect local interactions, and any deformation of the domain wall. It is nevertheless remarkable that just the combined effect of dipolar interactions and disorder yields power
law avalanche distribution in a multi-domain model. The latter appear, however, essential
to recover quantitatively the experimental results in bulk samples 
\cite{dz}. To completely resolve this issue, it would be necessary to study the dynamics of 
an array of flexible and interacting domain walls.

\section{Conclusions}

\label{conclusioni}

The technological and theoretical relevance of the hysteresis in ferromagnetic 
systems motivates the effort in modeling the phenomenon. 
Of particular interest is the
study of the dynamic hysteresis, connected with the power losses of
the material, and of the Barkhausen noise, which represents a tool for the
investigation of the magnetization dynamics.

In this article we have presented a model for the dynamics of a system of
parallel and rigid uncharged (Bloch) domain walls, moving in a disordered landscape driven by
an external magnetic field. The analyzed configuration of domain walls 
is a common case both in two and three dimensional samples. Our model is based
on the interplay between the long-ranged dipolar interactions and 
the external field and disorder contributions. Due to the simplicity of the
wall configuration, we calculated perturbatively the demagnetizing
factor and the magnetostatic susceptibility as functions of the geometrical
parameters of the system, namely the thickness, the height and the number of
domains at equilibrium, in the absence of external magnetic field. The
simulation results perfectly agree with the calculations, and we have thus been able 
to link a measurable macroscopic quantity to the microscopical system behavior.
Next, we have investigated the dynamic hysteresis by means of the study of the
coercive field behavior, obtained by simulating the equations of motion of the
walls array. The coercive field displays a power law behavior as a function
of the applied field rate, with an exponent in agreement with experimental
available data.  Finally we have analyzed the Barkhausen noise, constructing
the probability distributions of the Barkhausen avalanches size and time 
duration, and found power laws with cutoffs for both the distributions.
 Moreover we find that the probability distribution of domain walls velocity 
deviates at low driving rates from the ABBM related formula. Since the latter
model is based on the motion of a single domain wall, we can conclude that the 
multi-domain effects are important in ferromagnetic systems, since they affect
system parameters like the demagnetizing factor and the behavior of the 
Barkhausen noise.

\section{appendix}

We can develop the magnetostatic force $F_m(k)$ 
on a generic wall $k$ (see Eq. \ref{Fmk}), around its
equilibrium position:
$$F_m(k)=F_m^{eq}(k)+\delta F_m(k)=\delta F_m(k),$$
since the equilibrium term $F_m^{eq}(k)$ is $0$ by definition.
It follows:
$$F_m(k)=\sum_{j=0}^{n}\left.\frac{\partial F_m(k)}{\partial x_j}\right|_{u_j=0}(-1)^{j+1}u_j$$

\begin{equation}
=\sum_{j=1}^{n-1}\left.\frac{\partial F_m(k)}{\partial x_j}\right|_{u_j=0}(-1)^{j+1}u_j,
\label{sviluppo}
\end{equation}
where the last equality descends from the fact that the boundaries of the
sample are fixed.

For simplicity in the calculations, let us separate the contribution $k=j$
from the others, and rewrite Eq. \ref{sviluppo} as:

$$F_m(k)=\frac{\partial F_m(k)}{\partial x_k}u_k(-1)^{k+1}+\sum_{j\neq k}\left.\frac{\partial F_m(k)}{\partial x_j}\right|_{u_j=0}(-1)^{j+1}u_j$$

$$=2M_s^2\mu_0(-1)^{k+1}u_k\sum_{j=1}^{n-1}\left[\left(\frac{\partial^2 g(x_{j+1},x_k)}{\partial x_k^2}-\epsilon^2\frac{\partial^2g(x_{j},x_k)}{\partial x_k^2}\right)\right.$$

$$\left.-\epsilon^2 \left(\frac{\partial^2 f(x_{j+1},x_k)}{\partial x_k^2}-\epsilon^2\frac{\partial^2f(x_{j},x_k)}{\partial x_k^2}\right)\right]$$

$$+4M_s^2\sum_{j=0}^{n-1}(-1)^{j+k}\left(\frac{\partial^2 g(x_{j},x_k)}{\partial x_k^2}-\epsilon^2\frac{\partial^2f(x_{j},x_k)}{\partial x_k^2}\right)u_j(-1)^{j+1},$$
where we have used the symmetry properties of the derivatives 
(see Eq. \ref{derivate} ).
In the last sum we have included again the $k$-th term since it is null. 
If we neglect the contribution from the boundaries, that could be done since 
we are in the $n\rightarrow \infty$ case, we can translate $j+1\rightarrow j$
in the first sum, together with the sum index. We obtain:
$$
F_m(k)=4\mu_0M_s^2u_k\sum_j(-1)^j\left(\frac{\partial^2g(x_j,x_k)}{\partial x_k^2}-\epsilon^2\frac{\partial^2f(x_j,x_k)}{\partial x_k^2}\right)
$$

$$-4\mu_0M_s(-1)^k\sum_j(-1)^ju_j\left(\frac{\partial^2g(x_j,x_k)}{\partial x_k^2}-\epsilon^2\frac{\partial^2f(x_j,x_k)}{\partial x_k^2}\right).
$$

So finally, it results

$$F_m(k)=4\mu_0M_s^2\sum_{j=0}^{n-1}\left(\frac{\partial^2 g(x_j,x_k)}{\partial x_k^2}-\epsilon^2\frac{\partial^2f(x_j,x_k)}{\partial x_k^2}\right)$$

\begin{equation}
\times\left((-1)^ju_k-(-1)^ku_j\right),
\label{Fk_sviluppata}
\end{equation}
where the double derivatives of Eq. \ref{derivate} are:

\begin{equation}
\left\{
\begin{array}{l}
\displaystyle
\frac{\partial^2f(x,y)}{\partial x^2}=-\frac{2}{\sqrt{4d^2+(x-y)^2}}
\\\\
\displaystyle
\frac{\partial^2g(x,y)}{\partial x^2}=4\left(\sqrt{\epsilon^2+(x-y)^2-|x-y|}\right)
\\\\
\displaystyle
+2\epsilon\ln\left(\frac{\sqrt{\epsilon^2+(x-y)^2}-\epsilon}{\sqrt{\epsilon^2+(x-y)^2}+\epsilon}\right)
\end{array}
\right.
\label{derivate_seconde}
\end{equation}

In order to handle analytically Eq. \ref{Fk_sviluppata}, let us restrict
our study to the limit $\epsilon <<|x-y|<<L<<d$, which is the physical case
in most ferromagnetic samples. The second derivatives of Eq. 
\ref{derivate_seconde} thus become

\begin{equation}
\left\{
\begin{array}{l}
\displaystyle
\frac{\partial^2f(x,y)}{\partial x^2}=-\frac{1}{d}
\\\\
\displaystyle
\frac{\partial^2g(x,y)}{\partial x^2}=-\frac{2\epsilon^2}{|x-y|}
\end{array}
\right.
\label{derivate_seconde_limite}
\end{equation}
and $F_m(k)$ could be in this limit rewritten:
$$F_m(k)=4\mu_0M_s^2\sum_j\left(-\frac{2\epsilon^2}{|x_j-x_k|}+\frac{\epsilon^2}{d}\right)$$

\begin{equation}
\times(-1)^{j+k}\left((-1)^ku_k-(-1)^ju_j\right).
\label{Fk_lim}
\end{equation}

To have some insight on Eq. \ref{Fk_lim}, we can study the effect
of the magnetostatic force in two simplified situations:
\begin{itemize}
\item Let us imagine that all the walls are fixed but the $k$-th. Thus Eq. 
\ref{Fk_lim} can be rewritten:
$$F_m(k)=-8\mu_0M_s^2\epsilon^2\sum_i\frac{(-1)^{i+k}}{|x_i-x_k|}(-1)^ku_k$$
$$=-8\mu_0M_s^2\epsilon^2u_k\sum_i\frac{(-1)^{i}}{|x_i-x_k|},$$
that means that every wall $i$ contributes with a term of sign $(-1)^{i+1}$. 
Since the most important contribution in the sum of Eq. \ref{Fk_lim} is due to the
nearest neighbors, if $k$ is odd (even) and has thus a positive (negative)
perturbation on the position, it will be subject to a negative (positive)
magnetostatic force, recalling the wall to the equilibrium position. 
\item Considering the case with all the perturbations $u_i=u$, $\forall i$.
The magnetostatic force becomes
$$F_m(k)=4M_s^2\epsilon^2u\sum_i\left(-\frac{2}{|x_i-x_k|}+\frac{1}{d}\right)$$
\begin{equation}
\times(-1)^{i+k}\left((-1)^k-(-1)^i\right).
\label{fk}
\end{equation}
So only the walls with the opposite parity with respect to $k$ will contribute
to $F_m(k)$. Thus the even (odd) walls, that have a negative (positive)
displacement, will experience a positive (negative) force, recalling the wall 
to the equilibrium position. 
\end{itemize}

Let us restrict ourselves in the following to the limit above,  
$u_i=u$, $\forall i$. So, starting from Eq. \ref{fk}, we can write

$$F_m(k)=\frac{4\mu_0M_s^2\epsilon^2u}{d}\sum_i\left((-1)^i-(-1)^k\right)$$
$$-\frac{8\mu_0M_s^2\epsilon^2un}{L}\sum_i\frac{1}{|i-k|}\left((-1)^i-(-1)^k\right),$$
using the definition $x_i=iL/n$.
Since $L<<d$ and $n/|i-k|\geq 1$ $\forall$ $i,k$, we can approximate
$$F_m(k)\simeq -\frac{8\mu_0M_s^2\epsilon^2un}{L}\sum_i\frac{1}{|i-k|}\left((-1)^i-(-1)^k\right)$$
$$=\alpha \sum_i\frac{1}{|i-k|}\left((-1)^i-(-1)^k\right),$$
defining $\alpha=8\mu_0M_s^2\epsilon^2un/L$. In the sum
$$F_m(k)=\alpha (-1)^k\sum_i\frac{1}{|i-k|}\left((-1)^{i+k}-1\right),$$
the only non null terms are the $i$th terms $i'$ 
with opposite parity with respect to
$k$, that lead to a contribution
$$F_m(k)=\alpha(-1)^k(-2)\sum_{i'=0}^{n}\frac{1}{|i-k|}.$$
Since $n\rightarrow \infty$, we could consider $k$ in the middle of
the sample and write
$$F_m(k)=-4\alpha(-1)^k\sum_{i'=1}^{n/2}\frac{1}{i}=-4\alpha(-1)^k\sum_{i=1}^{n/4}\frac{1}{2i-1}$$
$$=-2\alpha(-1)^k\left[\gamma+\ln\left(\frac{n}{4}\right)+2\ln{(2)}\right],$$
by using the sum of an harmonic finite series of odd terms, where
$\gamma=0.577215$. So finally
$$F_m(k)=\frac{8\mu_0M_s^2\epsilon^2un}{L}(-1)^kA(n)$$
where
$$A(n)=2\left[\gamma+\ln\left(\frac{n}{4}\right)+2\ln{(2)}\right]=2\left[\gamma+\ln(n)\right].$$

The energy term due to the external magnetic field is given by 
Eq. \ref{intext}, and in this case could be written
$$E_{ext}=-4\mu_0M_sH_{ext}d\epsilon \sum_i u_i.$$
Therefore, $F_{ext}(k)$ is given by:
$$F_{ext}(k)=-\frac{\partial E_{ext}}{\partial x_k}=\frac{4\mu_0 M_sH_{ext}d\epsilon\sum_i u_i}{(-1)^{k+1}u_k}$$
$$=(-1)^{k+1}4 \mu_0M_s H_{ext}d\epsilon.$$
Since at the equilibrium the total force $F=0$, in absence of disorder it must 
be
$$F_m(k)+F_{ext}(k)=0$$
$$=\frac{8\mu_0M_s^2\epsilon^2un}{L}(-1)^kA(n)+(-1)^{k+1}4 \mu_0M_s H_{ext}d\epsilon,$$
from which follows
$$H_{ext}=\frac{2\epsilon u A(n)M_s n}{Ld}.$$
The magnetization $M$ could be written
$$M=\frac{4ud\epsilon M_s n}{2dL\epsilon}=\frac{2u M_s n}{L}.$$

So finally we can calculate the magnetic susceptivity $\chi$:
\begin{equation}
\chi=\frac{\partial M}{\partial H_{ext}}=\frac{M}{H_{ext}}=\frac{d}{\epsilon A(n)},
\label{chi_appendice}
\end{equation}
which was our goal.


\begin{thebibliography}{99}


\bibitem{bertotti}
G. Bertotti, {\it Hysteresis in magnetism} (Academic Press, San Diego, 1998)
\bibitem{RFIM}
J. P. Sethna, K. Dahmen, S. Kartha, J. A. Krumhansl, B. W. Roberts
and J. D. Shore, {\it Phys. Rev. Lett.} {\bf 70}, 3347 (1993)

\bibitem{sethna}
J. P. Sethna, K. A. Dahmen and C. R. Myers, 
{\it Nature}, {\bf 410} 242 (2001)

\bibitem{vives}
E. Vives and A. Planes, {\it Phys. Rev. B} {\bf 50} 3839 (1994)

\bibitem{vives2}
E. Vives and A. Planes, {\it J. Mag. Magn. Mat.} {\bf 221} 164 (2000)

\bibitem{vives3}
E. Vives and A. Planes, {\it Phys. Rev. B} {\bf 63} 134431 (2001)

\bibitem{chakrabarti}
B. K. Chakrabarti and M. Acharyya, 
{\it Rev. Mod. Phys.} {\bf 71}, 847 (1999)

\bibitem{acharyya}
M. Acharyya and B. K. Chakrabarti, {\it Phys. Rev. B} {\bf 52},
6550 (1995)

\bibitem{sides}
S. W. Sides, P. A. Rikvold and M. A. Novotny,
{\it Phys. Rev. E} {\bf 59},2710 (1999)

\bibitem{rikvold}
H. L. Richards, M. A. Novotny and P. A. Rikvold,
{\it Phys. Rev B} {\bf 54}, 4113 (1996)

\bibitem{rikvold1}
S. W. Sides, P. A. Rikvold and M. A. Novotny,
{\it Phys. Rev. E} {\bf 57}, 6512 (1998)

\bibitem{lyuksyutov}
I. F. Lyuksyutov, T. Nattermann and V. Pokrovsky, {\it Phys. Rev. B} {\bf 59}, 4260 (1999)

\bibitem{ABBM}
B. Alessandro, C. Beatrice, G. Bertotti and A. Montorsi,
{\it J. Appl. Phys.} {\bf 68} 2901 (1990)

\bibitem{zapperi}
S. Zapperi, P. Cizeau, G. Durin and H. E. Stanley,
{\it Phys. Rev. B} {\bf 58}, 6353 (1998)

\bibitem{zapperi2}
P. Cizeau, S. Zapperi, G. Durin and H. E. Stanley, 
{\it Phys. Rev. Lett.} {\bf 79} 4669 (1997)

\bibitem{urbach}
J. S. Urbach, R. C. Madison, J. T. Markert, 
{\it Phys. Rev. Lett.} {\bf 75} 276 (1995)

\bibitem{narayan}
O. Narayan, {\it Phys. Rev. Lett.} {\bf 77} 3855 (1996)

\bibitem{bahiana}
M. Bahiana, B. Koiller, S. L. A. de Queiroz, J. C. Denardin and R. L. Sommer,
{\it Phys. Rev. E} {\bf 59} 3884 (1999)

\bibitem{queiroz}
S. L. A. de Queiroz and M. Bahiana, {\it Phys. Rev. E} {\bf 64} 066127 (2001)

\bibitem{moore}
T. A. Moore and J. A. C. Bland, {\it J. Phys.: Condens. Matter} {\bf 16}, 
R1369 (2004)

\bibitem{nistor}
C. Nistor, E. Faraggi, and J. L. Erskine,
{\it Phys. Rev. B} {\bf 72}, 014404 (2005)

\bibitem{raquet}
B. Raquet, R. Mamy and J. C. Ousset, {\it Phys. Rev. B} {\bf 54} 4128 (1996)

\bibitem{he}
Y.-L. He and G.-C. Wang, {\it Phys. Rev. Lett.} {\bf 70} 2336 (1993)

\bibitem{luse}
C. N. Luse and A. Zangwill, {\it Phys. Rev. E} {\bf 50}, 224 (1994)

\bibitem{jiang}
Q. Jiang, H.-N. Yang and G.-C. Wang, {\it Phys. Rev. B} {\bf 52} 14911 (1995)

\bibitem{suen}
J.-S. Suen and J. L. Erskine, {\it Phys. Rev. Lett.} {\bf 78}, 3567 (1997)

\bibitem{bland1}
W. Y. Lee, B.-Ch. Choi, Y. B. Xu and J. A. C. Bland,
{\it Phys. Rev. B} {\bf 60}, 10216 (1999)

\bibitem{choi}
B. C. Choi, W. Y. Lee, A. Samad and J. A. C. Bland,
{\it Phys. Rev. B} {\bf 60}, 11906 (1999)

\bibitem{lee2}
W. Y. Lee, Y. B. Xu, S. M. Gardiner and J. A. C. Bland,
{\it J. Appl. Phys.} {\bf 87}, 5926 (2000)

\bibitem{lee3}
W. Y. Lee, A. Samad, T. A. Moore and J. A. C. Bland and B. C. Choi,
{\it Phys. Rev. B} {\bf 61}, 6811 (2000)

\bibitem{suen2}
J.-S. Suen, M. H. Lee, G. Teeter and J. L. Erskine, {\it Phys. Rev. B} {\bf 59}, 4249 (1999)

\bibitem{santi}
L. Santi, R. Sommer, A. Magni, G. Durin, F. Colaiori and S. Zapperi,
{\it IEEE Trans. Magn.} {\bf 39}, 2666 (2003) 

\bibitem{benedetta}
B. Cerruti and S. Zapperi, {\it Phys. Rev. B},
{\bf 75}, 064416 (2007)

\bibitem{colaiori}
F. Colaiori, G. Durin and S. Zapperi,
{\it Phys. Rev. Lett.}{ \bf 97}, 257203 (2006).

\bibitem{durin06}
G. Durin and S. Zapperi, {\it J. Stat. Mech.}, P01002, (2006)

\bibitem{barkhausen}
H. Barkhausen,
{\it Z. Phys.} {\bf 20} 401 (1919)

\bibitem{BK}
G. Durin and S. Zapperi, 
{\it The Science of Hysteresis: Physical Modeling, Micromagnetics, and Magnetization Dynamics} vol II (Amsterdam: Academic) chapter III (The Barkhausen Noise) pp 181-267 [cond-mat/0404512] (2005) 

\bibitem{wisho}
H. J. Williams and W. Shockley, 
{\it Phys. Rev.} {\bf 75} 178 (1949)

\bibitem{dz}
G. Durin and S. Zapperi, 
{\it Phys. Rev. Lett.} {\bf 84} 4705 (2000)

\bibitem{kovacs}
K. Kov\'acs, Y. Brechet and Z. N\'eda,
{\it Modelling Simul. Mater. Sci. Eng.} {\bf 13} 1341 (2005)

\bibitem{hubert}
A. Hubert and R. Schaefer, 
{\it Magnetic domains} (Springer: New York ) (1998)

\bibitem{walsh}
B. Walsh, S. Austvold and R. Proksch, 
{\it J. Appl. Phys.} {\bf 84} 5709 (1998)

\bibitem{schwarz}
A. Schwarz, M. Liebmann, U. Kaiser, R. Wiesendanger, T. W. Noh  and D. W. Kim,
{\it Phys. Rev. Lett.} {\bf 92} 077206 (2004)

\bibitem{kim}
D. -H. Kim, S. -B. Choe and  S. -C. Shin, 
{\it Phys. Rev. Lett.} {\bf 90} 087203 (2003)

\end{thebibliography}
\end{document}